# KINEMATICS OF ELECTRONS IN THE VOLUME OF A PLANAR VACUUM DIODE IN REGIME OF SATURATION. PARAMETERS OF HYSTERESIS

**Dimitar G. Stoyanov**

Faculty of Engineering and Pedagogy in Sliven, Technical University of Sofia
59, Bourgasko Shaussee Blvd, 8800 Sliven, BULGARIA
e-mail: dgstoyanov@abv.bg

**Abstract** *The kinematics laws of electrons motion in the volume of a planar vacuum diode running in regime of current saturation are used. The characteristics of diode hysteresis in the conditions of S-figurative instability are got and analyzed.*
**Key words** *a planar vacuum diode in regime of saturation, electrons velocity at anode*

## 1. Introduction

Regardless of the fact that more than 100 years have been passed since the first publications concerning the theory of vacuum thermo-electron diode [1] occur, the problems of that kind of diode are a question of present interest because of the wide applications of elements of that theory in practice and technologies [2-4]. The investigation target is the physical processes taking place in ensembles of similar particles, for example electrons. During the studying of physical processes running in similar particles ensembles one-partial approximation may be applied in a series of cases. The motion of a single particle that is moving in the equivalent total field created by the interaction of the considered particle with all other ensemble particles is studied.

The use of the kinematics method of approach [5] enables studying and analyzing of the charged particles motion in vacuum diode volume. The obtained solutions throw light on processes which cannot be analyzed in detail by the potential method of approach [4]. It is especially highly estimated during the analysis of S-figurative instability in diode current-voltage characteristic which generates functional hysteresis [4].

The **goal** of this article is the kinematics solutions described in [5] to be analyzed as well as the kinematics of charged particles in the planar vacuum diode volume to be investigated in regime of current saturation. The investigation has to show up the characteristics of the S-figurative instability caused by the solution polysemy.

## 2. Kinematics of electron
### 2.1 Theoretical formulation and general solution

The planar vacuum diode [4, 5] is a system of two in parallel metallic plain electrodes with surface **S**, situated apart in a distance **d.** Electrons with electrical charge **q (q<0**) are emitted from one of the electrodes, i.e. the cathode. In our analysis the electrons initial velocity $V_0$ is equal, and the electrons possess density of the emitted current $j_0$ [4]. We accept that the cathode is situated on the plane YOZ of a Cartesian coordinate system and has a coordinate **x = 0**. Thus, the anode will have a coordinate **x = d**. We suppose that **d** is significantly great, and is always greater than $x_m$. The direction of the initial velocity vector is perpendicular to cathode plain. On such geometry of the problem all physical vectors are in parallel to the axis OX, and all quantities will depend only on the coordinate **x**. For the simplification of the equations recordings the quantities will be non-dimensional.

As a scale of the velocity, the coordinate $x$, and the time we will use $V_0$, the distance between the electrodes $d$, and the magnitude $d/V_0$, respectively.

The kinematics method of approach for the investigation of the charged particles motion in a planar vacuum diode volume is applied in [5]. After the scaling of quantities is done, we can write [5]:

$$\frac{d^3x}{dt^3} = \frac{j \cdot q}{\varepsilon_0 \cdot m} \cdot \frac{d^2}{V_0^3} = b = \text{const.} \qquad (1)$$

In [5] is also shown that at constant $b$ only the solutions with not-negative values of $V_m$ have a physical meaning. If $V_m > 0$, then all electrons emitted by the cathode reach the anode and a current with density $j_0$ will flow through the diode. This current does not depend on the applied anode voltage. Under the existing regime the constant $b$ will have a value, as following

$$b_0 = \frac{q}{\varepsilon_0 \cdot m} \cdot \frac{d^2}{V_0^3} \cdot j_0 \qquad (2)$$

This dependence has the form of a horizontal straight line on the current-voltage characteristic of the vacuum diode. This is ***the regime of current saturation through the vacuum diode.***

The solution of $x(t)$ from eq. (1) is searched by stage integration. Time will be read from the moment of electron emission. After the first integration of eq. (1) the electron acceleration is got:

$$a(t) = \frac{d^2x}{dt^2} = a_0 + b_0 \cdot t = b_0 \cdot (t - t_m). \qquad (3)$$

Here in eq. (3) with $a_0$ is marked the initial acceleration (in non-dimensional units), and with $t_m$ is marked the time moment in which the acceleration is nullified. The initial acceleration $a_0$ has a particular value at each set anode voltage, but this value changes at each change of anodic voltage value.

The second integration of eq. (1) comes to electron velocity:

$$V(t) = \frac{dx}{dt} = V_0 + a_0 \cdot t + \frac{b_0 \cdot t^2}{2} = V_m + b_0 \cdot \frac{(t - t_m)^2}{2}. \qquad (4)$$

In eq. (4) with $V_m$ is marked the minimum electron velocity at the moment $t = t_m$.

$$V_m = V_0 - b_0 \cdot \frac{t_m^2}{2}. \qquad (5)$$

The third integration of eq. (1) comes to the law of electron motion:

$$x = V_0 \cdot t + \frac{a_0 \cdot t^2}{2} + \frac{b_0 \cdot t^3}{6} = x_m + V_m \cdot (t - t_m) + b_0 \cdot \frac{(t - t_m)^3}{6}. \qquad (6)$$

At the moment $t = t_m$ the electron will have a coordinate $x = x_m$, where:

$$x_m = V_m \cdot t_m + b_0 \cdot \frac{t_m^3}{6}. \qquad (7)$$

At the moment $t = t_d$ the electron falls onto the anode and will possess velocity $V_d$ and a coordinate $d$. We can express $V_d$ and $d$ by $t_d$ from eq. (4) and eq. (6) in the following forms:

$$V_d = V_0 + a_0 \cdot t_d + \frac{b_0 \cdot t_d^2}{2}. \qquad (8)$$

$$d = V_0 \cdot t_d + \frac{a_0 \cdot t_d^2}{2} + \frac{b_0 \cdot t_d^3}{6}. \qquad (9)$$

### 2.2 An analysis of the solution

The curve of the law of charged particles velocity eq. (4) as a function of the time at some values of $V_m$ is shown as an illustration in Fig. 1.

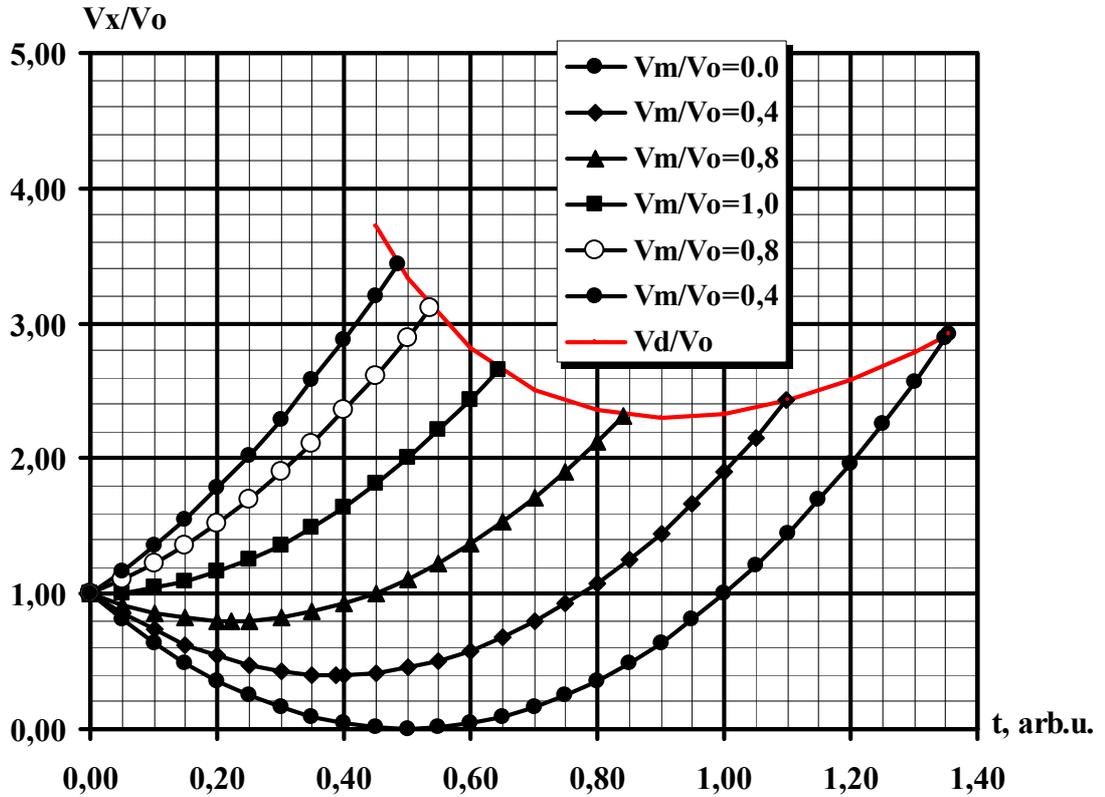

Fig. 1. Curves of the law of velocity eq. (4) at $b_0 = 8$, initial velocity $V_0 = 1$, and different $V_m$ (in non-dimensional units).

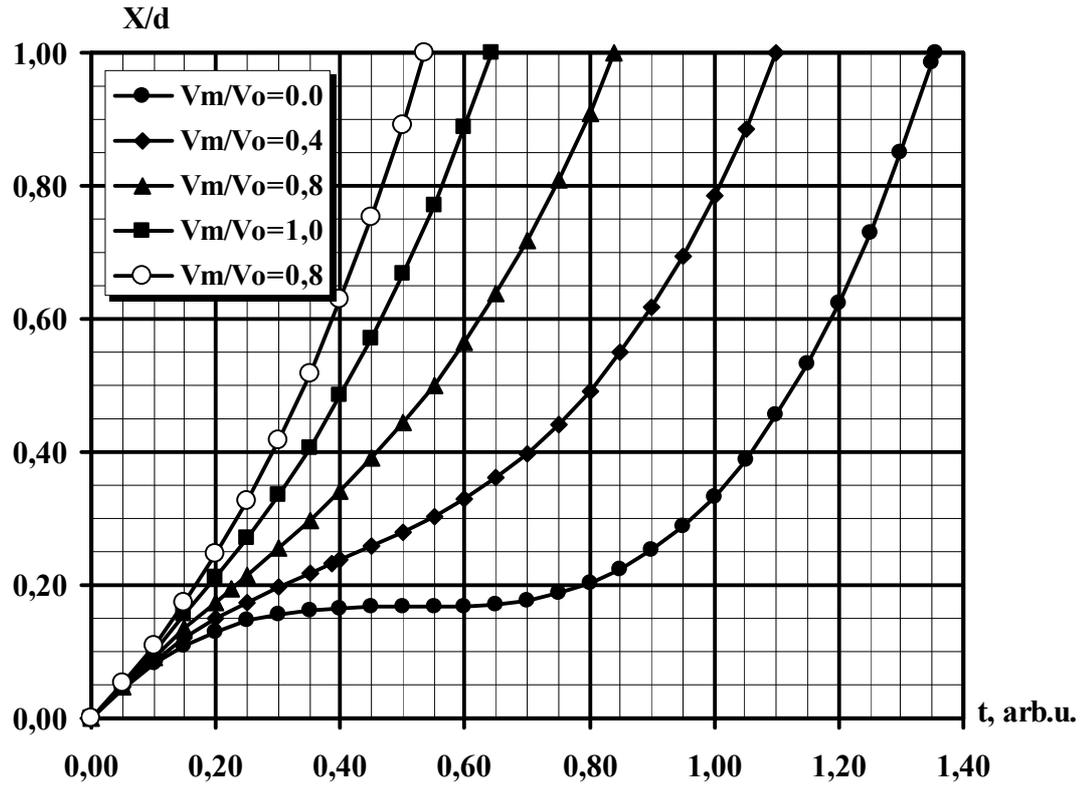

**Fig. 2.** Curves of the law of motion eq. (6) at $b_0 = 8$, initial velocity $V_0 = 1$, and different $V_m$ (in non-dimensional units).

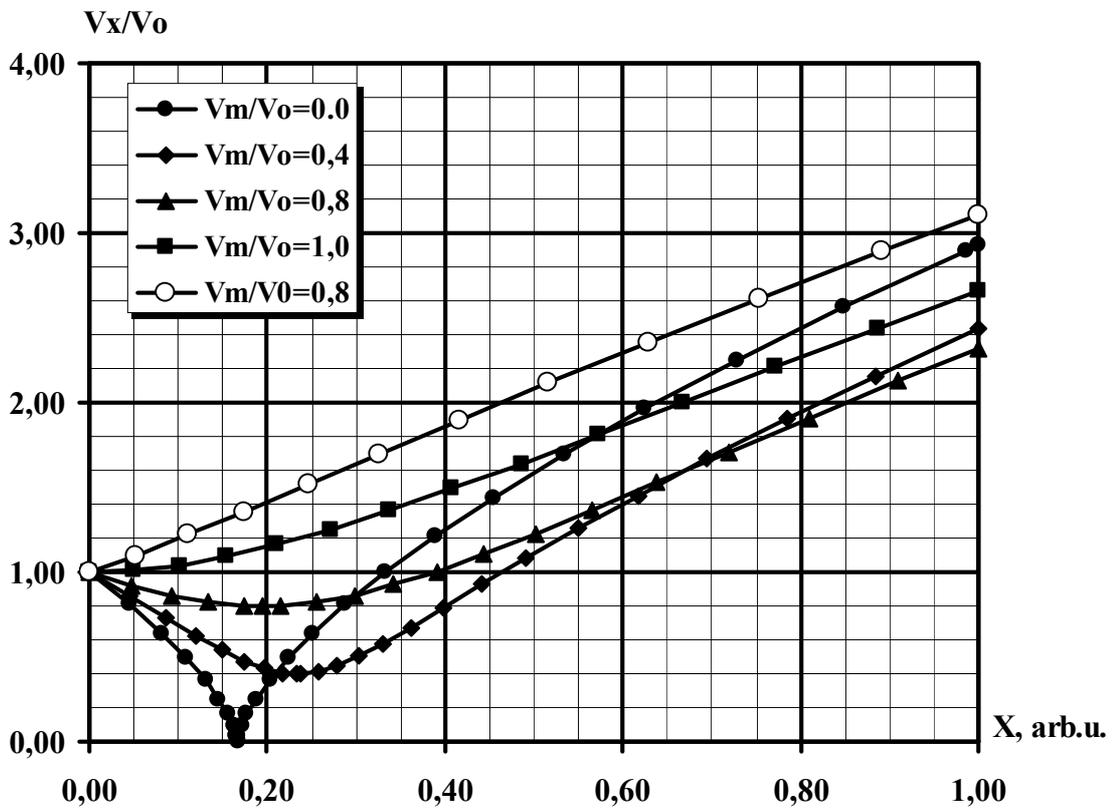

**Fig. 3.** Dependence of particles velocity on the coordinate x at $b_0 = 8$, and different $V_m$ (in non-dimensional units).

The time dependences of the coordinates of the charged particles (see eq. (6)) for some values of $V_m$ are shown in Fig. 2, while the graphic dependence of velocity on the charged particles coordinates for the same values of $V_m$ is given in Fig. 3. As it is evident from the curves the time dependences are monotonous, and there are no interruptions on the curves for the different values of $V_m$, whereas in Fig. 3 a tangle of interrupted curves is observed. These curves illustrate the presence of non-monotonous and non-synonymous dependences.

As a result of these curves we can observe that the minimum velocity $V_m$ cannot be higher than $V_0$. This fact could be explained in the following way.

From eq. (4) we can get

$$t_m = \pm \sqrt{\frac{2(V_0 - V_m)}{b}}. \qquad (10)$$

Thus, it is evident from eq. (10) that if $V_m$ is higher than $V_0$ then eq. (10) has no real solution. Besides, the negative values of $t_m$ are physically admissible. They correspond to working conditions of the vacuum diode at which the velocity curves (see Fig. 1) have no minimum inside the vacuum diode volume.

The time ($t_m$) dependences of $V_m$ and $V_d$ are shown in Fig. 4, while the time ($t_m$) dependences of $x_m$ and $t_d$ are shown in Fig. 5, as all dependences are got numerically. Finally, in Fig. 6 are given the graphic dependences of $x_m$, $t_d$, $V_m$ and $t_m$ on $V_d$, got also numerically.

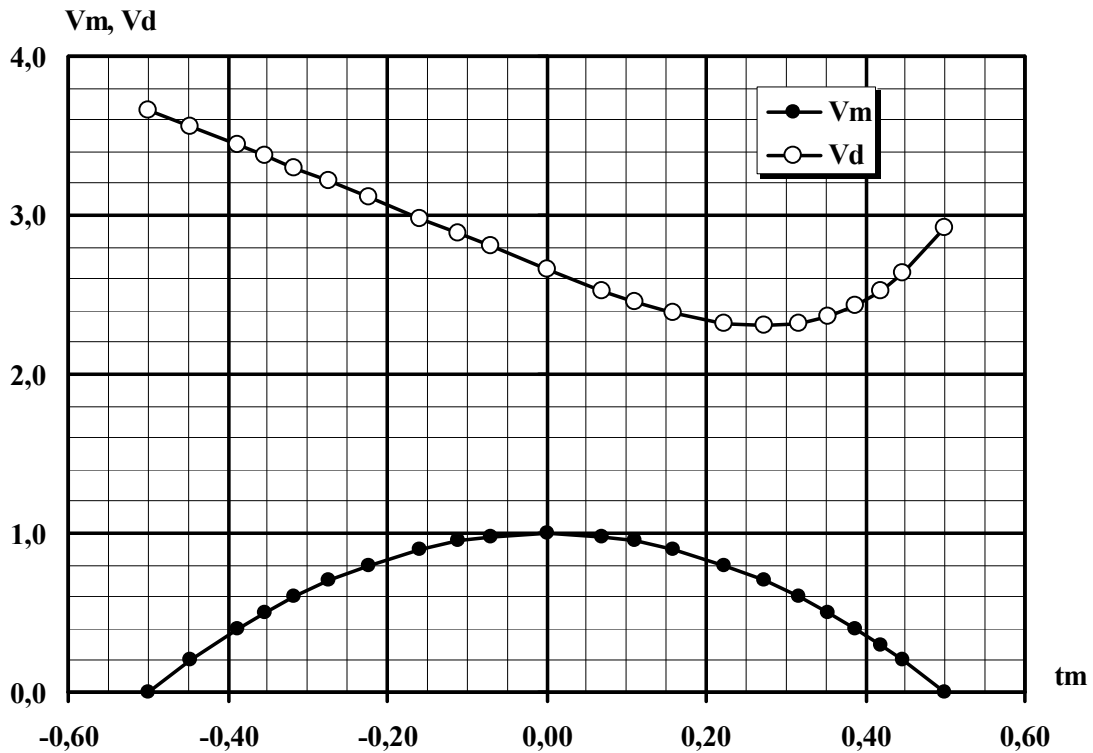

**Fig. 4. Time ($t_m$) dependences of $V_m$ and $V_d$ at $b_0 = 8$.**

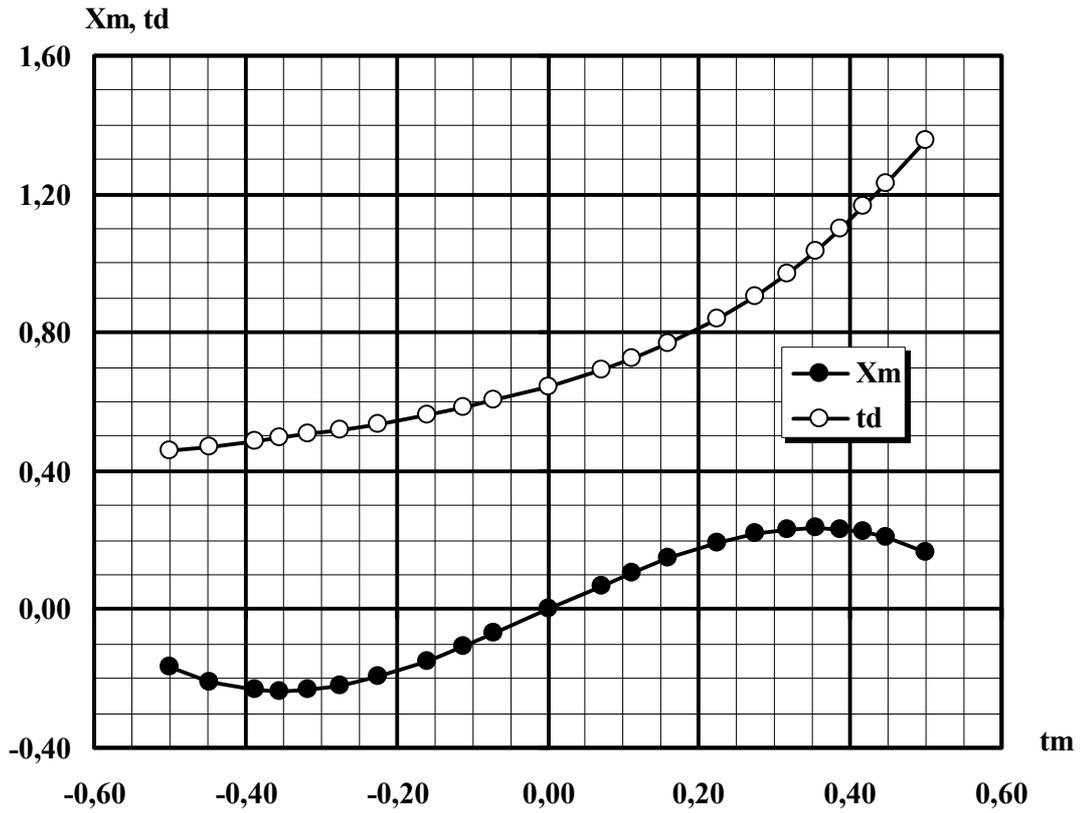

Fig. 5. Time ($t_m$) dependences of $x_m$ and $t_d$ at $b_0 = 8$.

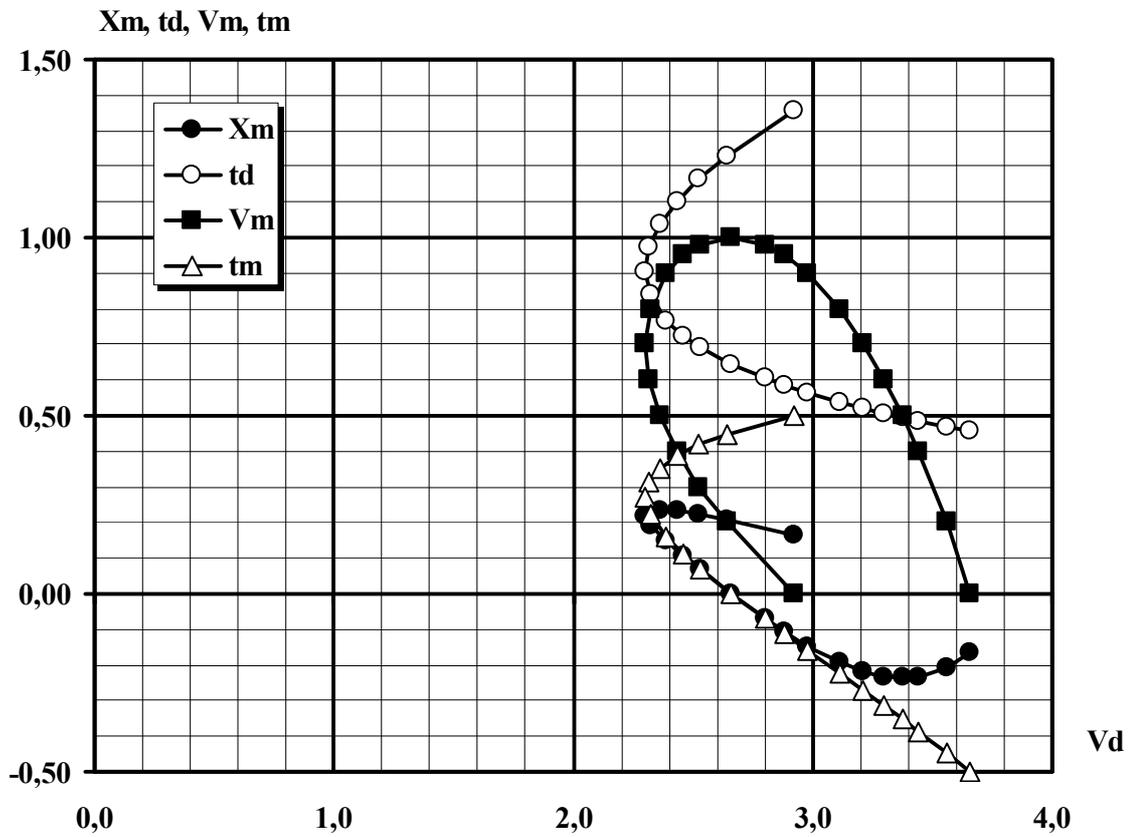

Fig. 6. Graphic dependences of $x_m$, $t_d$, $V_m$ and $t_m$ on $V_d$ at $b_0 = 8$.

From Fig.4 and Fig. 5 we observe the existence of solutions unambiguity in respect to $t_m$, while in Fig. 6 there is an ambiguity in respect to $V_d$, and this explains the availability of S-figurative instability [4] on the current-voltage characteristic of the planar vacuum diode. It is displayed in the presence of hysteresis during the oscillography of the current-voltage characteristic.

### 2.3 Parameters of hysteresis

The easy curve in the upper end of Fig. 1 corresponds to $V_d$ from eq. (8). The curve shows that the function is non-monotonous with minimum. The function dependence and the position of this minimum can be got analytically applying the following method of approach.

We consider eq. (8) and eq. (9) as a system of equations. From both equations we eliminate $a_0$ and get

$$V_d = -V_0 + \frac{2.d}{t_d} + \frac{b_0 . t_d^2}{6}. \qquad (11)$$

This equation represents the searched dependence of $V_d$ on $t_d$.

The value of $V_d$ in the minimum $V_{dm}$ we will get by the differentiation of eq. (11) in respect to $t_d$. The derivative will be nullified at $t_{dm}$, which will have the following value

$$t_{dm}^3 = \frac{6d}{b_0}. \qquad (12)$$

Then replacing in eq. (11) the value from eq. (12) we get

$$V_{dm} = -V_0 + \frac{3.d}{t_{dm}}. \qquad (13)$$

As an illustration of the authenticity of the obtained results, the graphic dependence of the law of particle velocity onto the anode $V_d$ from eq. (11) on the time necessary for the falling onto the anode $t_d$ at $b_0 = 8$ is shown in Fig. 7.

The curve at $b_0 = 0$ (zero current) describes the motion of a single charged particle in vacuum capacitor. The curve dependence is of the form:

$$V_d = -V_0 + \frac{2.d}{t_d}. \qquad (14)$$

The dependence from eq. (13) is shown as $V_{dm}$ in Fig. 7.

The points shown in Fig. 7 are a result of the calculations described in section 2.2 of the article. These points lie very well on the curve, describing eq. (11) at $b_0 = 8.$

The boundary point between the regime of the changing current and the regime of current saturation in the vacuum diode corresponds to $V_m = 0$. These are the rightest points in Fig. 1 and Fig. 7. According to [5] for these points we have

$$V_d^{3/2} + V_0^{3/2} = 3 \cdot \sqrt{\frac{b_0}{2}} \cdot d \qquad (15)$$

In order to get the analytical dependence of the coordinates of this point we will write eq. (4) for the cathode and anode, and after root extraction and summation by members we get

$$V_d^{1/2} + V_0^{1/2} = \sqrt{\frac{b_0}{2}} \cdot t_d \qquad (16)$$

If we consider eq. (15) and eq. (16) as a system of equations we can exclude $b_0$, and we get

$$\frac{V_d^{3/2} + V_0^{3/2}}{V_d^{1/2} + V_0^{1/2}} = \frac{3 \cdot d}{t_d} \qquad (17)$$

The dependence of eq. (17) is shown in Fig. 7 as a case when $V_m = 0$.

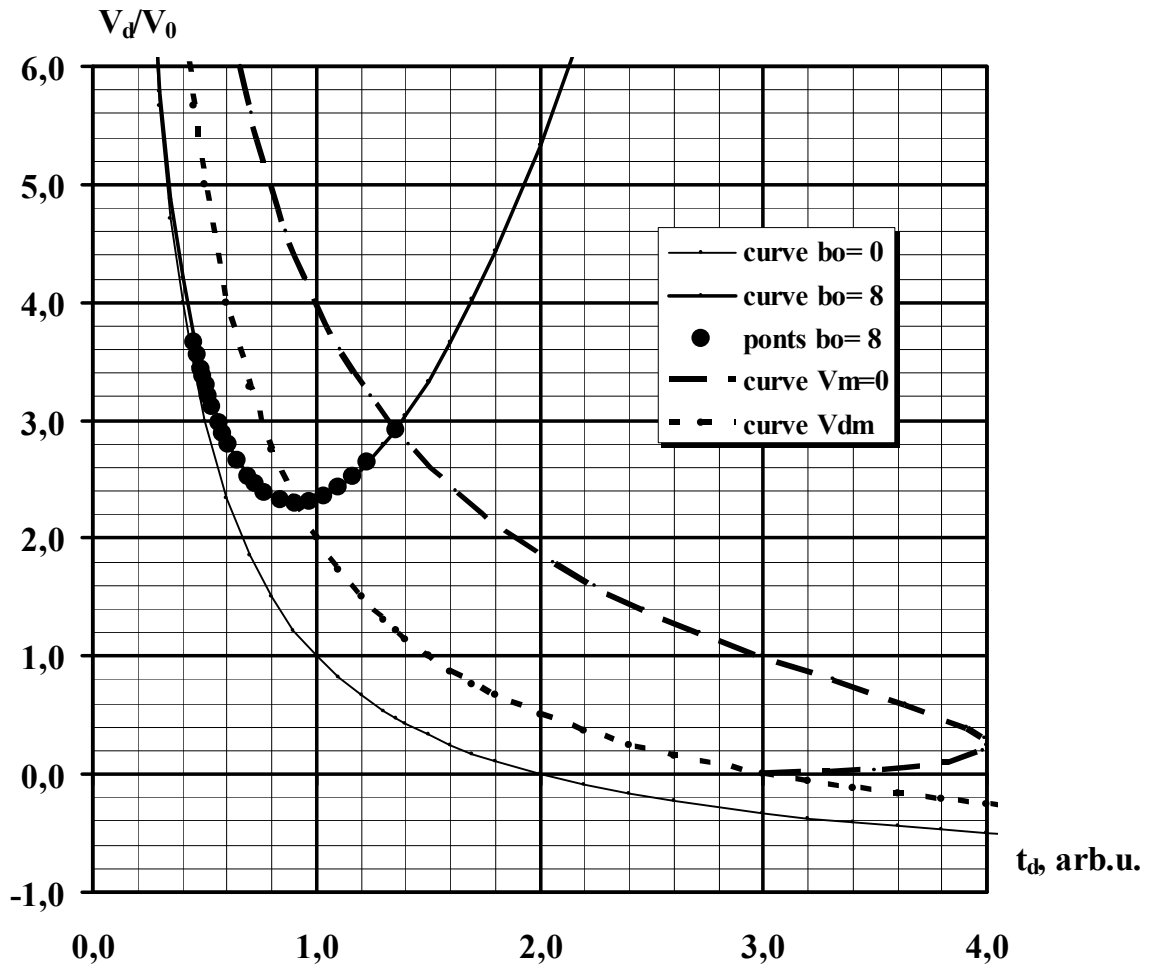

**Fig.7. Time ($t_d$) dependence of particles velocity on the anode at $b_0 = 0$ and $b_0 = 8$.**

From the contentions made in [4] follows that during the work of a vacuum diode in regime of current saturation, because of the presence of S-figurative instability, a hysteresis in the current-voltage characteristic of the diode is read. This hysteresis represents the following.

If we start the work of a vacuum diode at zero anode voltage and begin to increase the anode voltage, the velocity of electrons with which they will reach the anode will increase gradually as together with the current increase through the diode [5]. The diode works in a regime of current increase. When reaching the velocity that can be determined from eq. (15), the diode passes into regime of current saturation, and during the further increase of anode voltage the current will no longer change. Therefore, the electrons motion with velocity higher or equal to that determined from eq. (15) is a necessary and sufficient condition for the creation of the necessary current. Besides, eq. (15) can be transformed in the form:

$$b_0 = \frac{2}{9} \cdot \frac{\left(V_0^{3/2} + V_d^{3/2}\right)^2}{d^2} \tag{18}$$

Let us now start the work of diode in regime of current saturation. Then the electrons velocity will be higher than the value which may be determined from eq. (15). When we start to reduce the anode voltage the current will not change until the reach of the current velocity, determined from eq. (15). If we more decrease the voltage the diode will not switch over in regime of current increase. But when the anode voltage becomes lower than the necessary one for the formation of velocity $V_{dm}$, the diode will switch over the curve of regime of current change with current surge. Therefore the electrons motion with velocity higher or equal to $V_{dm}$ is a necessary and sufficient condition for the maintenance of necessary current. Besides, we can get the following dependence

$$b_0 = \frac{2}{9} \cdot \frac{\left(V_0 + V_{dm}\right)^3}{d^2} \tag{19}$$

### 3. Conclusions

The kinematics laws of electrons motion in the volume of a planar vacuum diode working in regime of current saturation are used. The parameters of a diode in the circumstances of S-figurative instability are got and analyzed. The minimum electrons velocity onto the anode necessary for the maintenance of the diode work in regime of current saturation is determined.